\documentclass[fleqn,10pt]{wlscirep}
\usepackage[utf8]{inputenc}
\usepackage[T1]{fontenc}
\usepackage{layouts}
\usepackage{setspace}
\usepackage{longtable}
\usepackage[version=4]{mhchem}       %Also for writing chemical equations
\usepackage{multirow}
\usepackage{array}
\usepackage{hyperref}
\usepackage[version=4]{mhchem}
\usepackage{float}
\usepackage{tikz}
\usepackage{xr}
\usepackage{cleveref}

\definecolor{light-gray}{rgb}{0.9,0.9,0.9}
\usepackage{listings}

\definecolor{light-gray}{rgb}{0.95,0.95,0.95}

\titleformat{\chapter}[hang] % To change the formatting of the chapter.
  {\large\bfseries}
  {}
  {0em}
  {}

\titlespacing*{\chapter} % Margin from left, up, and down.
  {0pt}    % left margin
  {-30pt}   % space above
  {10pt}   % space below

\newcolumntype{C}[1]{>{\centering\arraybackslash}p{#1}}

\newcolumntype{P}[1]{>{\centering\arraybackslash}p{#1}}

\title{Protocol for Clustering 4DSTEM Data for Phase Differentiation in Glasses}

\author[1]{Mridul Kumar}
\author[1, *]{Yevgeny Rakita}

\affil[1]{Department of Materials Engineering, Ben-Gurion University of the Negev, Be'er Sheva, Israel}

\affil[*]{Corresponding author email: rakita@bgu.ac.il}


\begin{abstract}

Phase-change materials (PCMs) such as Ge-Sb-Te alloys are widely used in non-volatile memory applications due to their rapid and reversible switching between amorphous and crystalline states. However, their functional properties are strongly governed by nanoscale variations in composition and structure, which are challenging to resolve using conventional techniques. Here, we apply unsupervised machine learning to 4-dimensional scanning transmission electron microscopy (4D-STEM) data to identify compositional and structural heterogeneity in Ge-Sb-Te. After preprocessing and dimensionality reduction with principal component analysis (PCA), cluster validation was performed with t-SNE and UMAP, followed by k-means clustering optimized through silhouette scoring. Four distinct clusters were identified which were mapped back to the diffraction data. Elemental intensity histograms revealed chemical signatures change across clusters, oxygen and germanium enrichment in Cluster 1, tellurium in Cluster 2, antimony in Cluster 3, and germanium again in Cluster 4. Furthermore, averaged diffraction patterns from these clusters confirmed structural variations. Together, these findings demonstrate that clustering analysis can provide a powerful framework for correlating local chemical and structural features in PCMs, offering deeper insights into their intrinsic heterogeneity.

\end{abstract}

\begin{document}

\flushbottom
\maketitle
\thispagestyle{empty}

%\noindent Please note: Abbreviations should be introduced at the first mention in the main text – no abbreviations lists. Suggested structure of main text (not enforced) is provided below.

\section*{Introduction}

Phase-change materials (PCMs), particularly those based on Ge-Sb-Te (GST) alloys, are foundational to non-volatile memory technologies such as phase-change random-access memory (PCRAM), neuromorphic computing, and optical data storage \cite{jones2024phase,lan2020thermophysical}. These materials exhibit rapid and reversible phase transitions between amorphous and crystalline states, which underpins their high-speed switching and non-volatility. The performance, endurance, and scalability of PCMs are intimately linked to nanoscale variations in local composition, chemical ordering, and structural arrangement. Even minor heterogeneities can profoundly impact crystallization kinetics, thermal stability, and electronic properties, highlighting the need for spatially resolved characterization at the atomic and nanoscale level.

Traditional characterization techniques, including X-ray diffraction (XRD) and conventional transmission electron microscopy (TEM), often provide only averaged or limited structural information, making it challenging to resolve subtle nanoscale heterogeneities \cite{ortega2022material,epp2016x,tang2017transmission}. To overcome these limitations, four-dimensional scanning transmission electron microscopy (4D-STEM) has emerged as a powerful approach. By recording a full diffraction pattern at every scan position, 4D-STEM generates a rich dataset containing both spatial and reciprocal-space information, capable of revealing local structural variations, strain, and lattice distortions \cite{Savitzky2021py4DSTEM,10.1017/S1431927619001351}. However, the very richness of these datasets poses a significant challenge. The resulting high-dimensional data are difficult to analyze using conventional methods, requiring advanced computational tools to extract meaningful insights.

Machine learning, particularly unsupervised learning, has recently shown great promise in analyzing complex materials datasets \cite{sadri2024unsupervised}. Dimensionality reduction techniques such as principal component analysis (PCA), t-distributed stochastic neighbour embedding (t-SNE), and uniform manifold approximation and projection (UMAP) can identify intrinsic patterns and reduce data complexity \cite{masonSelectingPrincipalComponents1985,maaten2008visualizing,mcinnes2018umap}. Clustering algorithms, including k-means and hierarchical clustering, can then group similar regions based on structural or compositional features, revealing hidden heterogeneity without prior assumptions \cite{ostrovsky2013effectiveness}. Recent studies have successfully applied these approaches to 4D-STEM and other diffraction datasets, uncovering subtle chemical variations, strain distributions, and nanoscale phase separations that were previously inaccessible \cite{allen2021fast, shi2022uncovering}.

In this work, we develop a comprehensive unsupervised clustering pipeline to analyze 4D-STEM diffraction data from multiple scans of same Ge-Sb-Te sample resulting in 6 files names 0033 - 0038. After preprocessing to remove artifacts, PCA was applied to reduce dimensionality, followed by cluster validation using t-SNE and UMAP. Clusters were assigned using k-means with silhouette scoring and projected back to the original data for visualization. Elemental intensity histograms and averaged diffraction patterns were used to characterize chemical and structural differences across clusters. Our analysis reveals distinct compositional signatures such as oxygen and germanium enrichment, tellurium and antimony variations and corresponding structural differences, demonstrating the power of clustering for resolving nanoscale heterogeneity in PCMs. This study not only provides new insights into the fundamental structure composition relationships in Ge-Sb-Te but also establishes a generalizable framework for applying machine learning to high-dimensional materials datasets, paving the way for the design of next-generation functional materials.

\section*{Results and Discussion}

\begin{figure}[t]
\centering
\includegraphics[width=12cm]{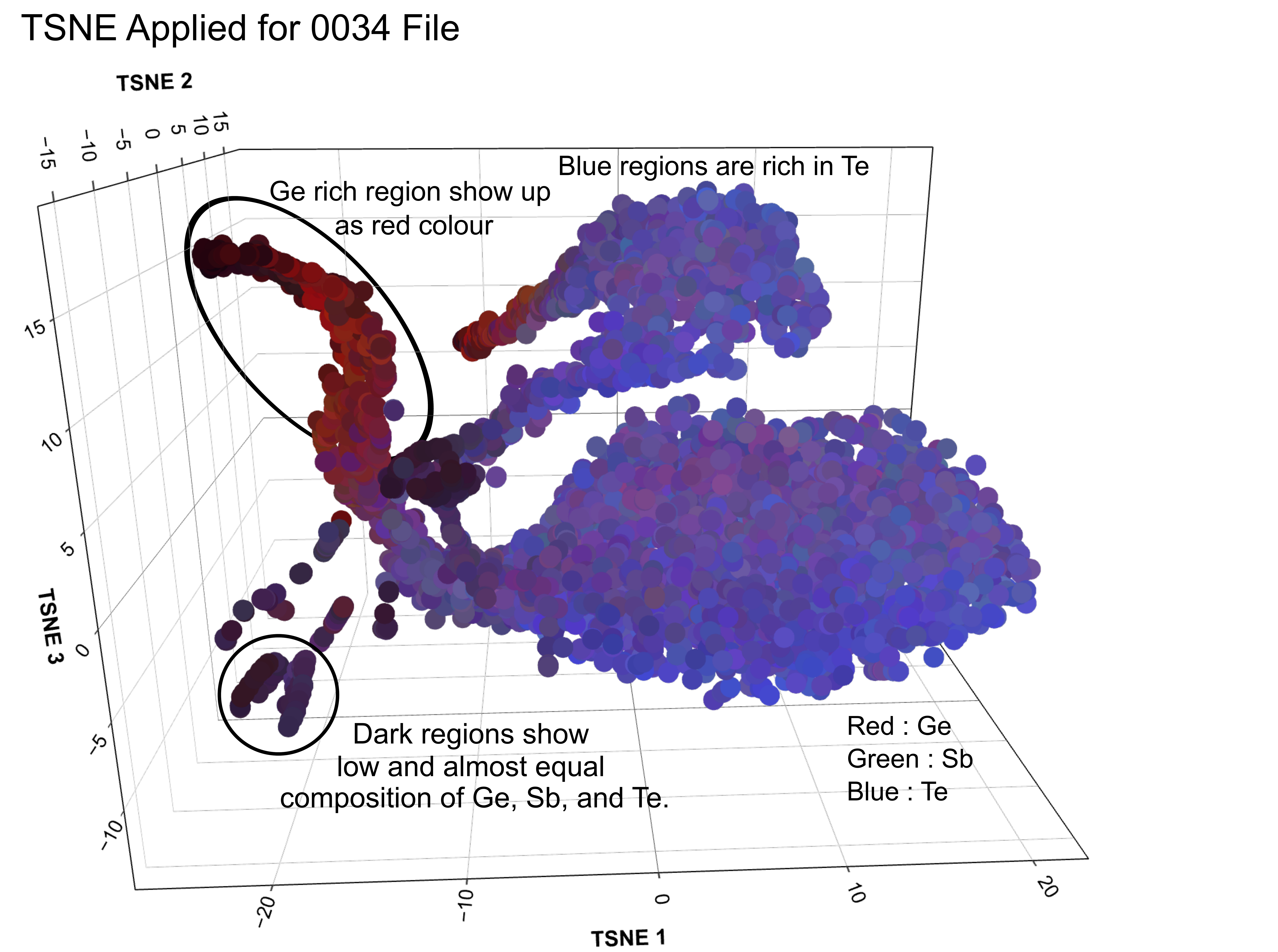}
\caption{This figure shows the application of t-SNE on reduced feature data of 4DSTEM file 0034. It is apparent from the figure that Ge rich regions cluster together at different location compared to the others.}
\label{fig:tsne}
\end{figure}

\begin{figure}
\centering
\includegraphics[width=\textwidth]{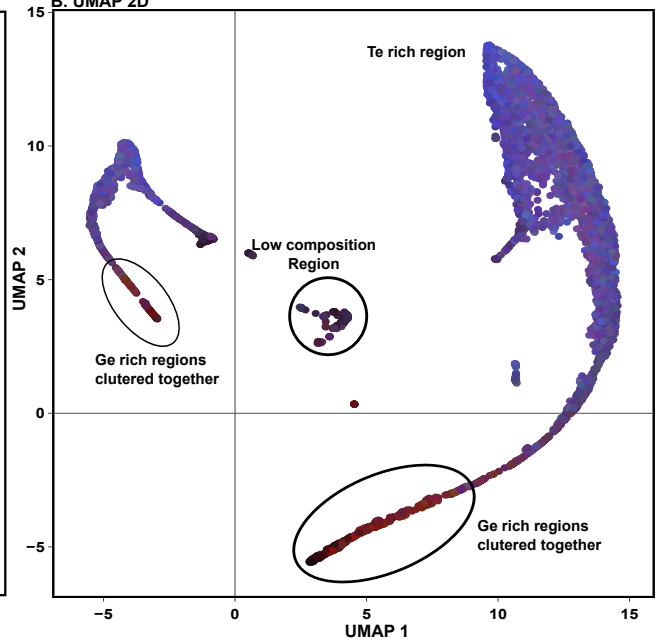}
\caption{This figure shows the application of UMAP on reduced data of 0034 file and the clustering that comes out of it. It is clear that the regions rich in Ge are now clustered much better compared to t-SNE. Here, A. Shows the 3D projection of the data and B. Shows 2D projection of the data.}
\label{fig:umap}
\end{figure}

\begin{figure}
\centering
\includegraphics[width=12cm]{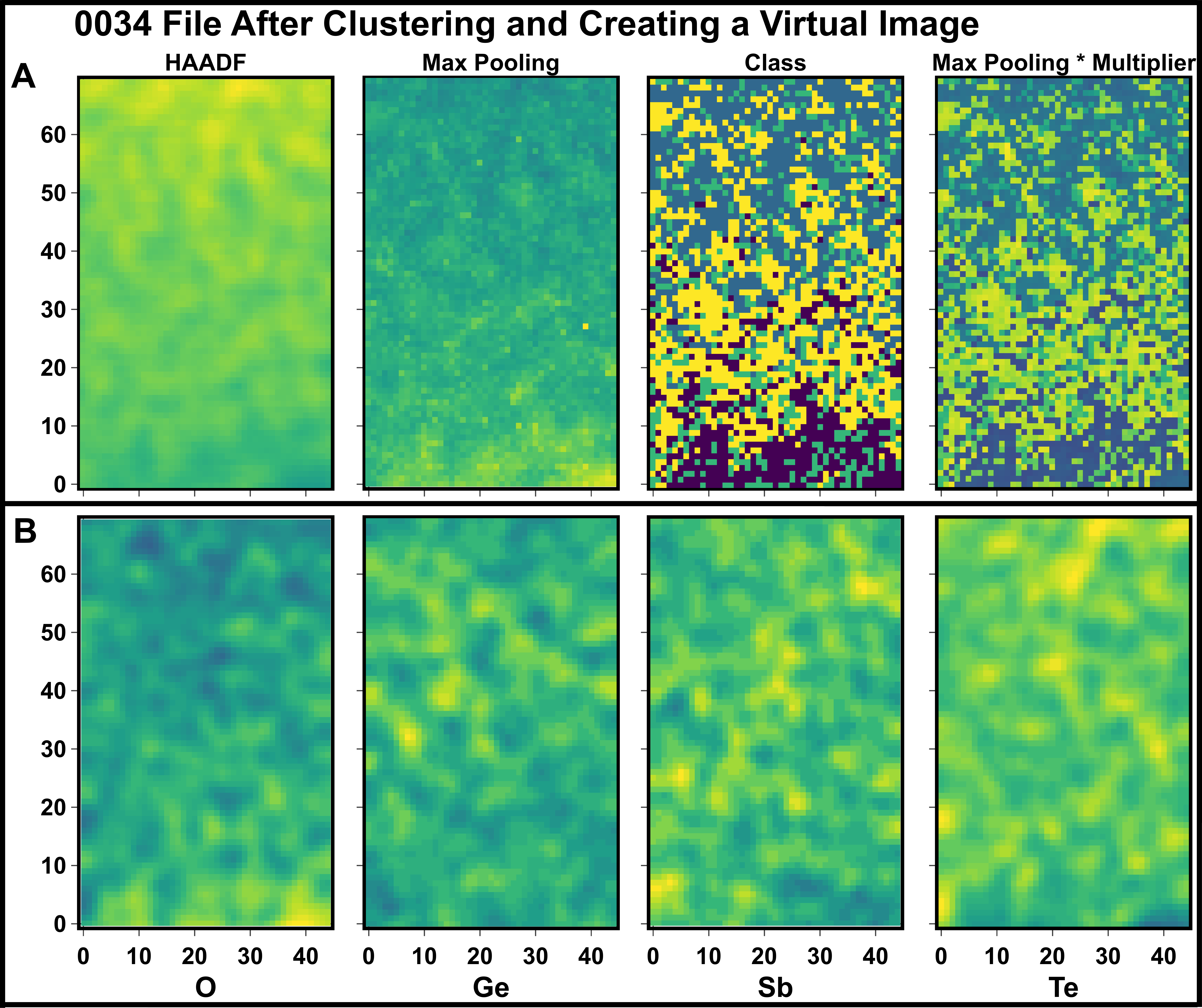}
\caption{This figure shows the segregation of the real space of the sample. A. HAADF of the sample with maxpooling for creating the virtual image. Projection of the clustering classes onto the sample after application of k-means clustering. B. It shows the EDS maps of the elements in file 0034.}
\label{fig:kmeans}
\end{figure}

\begin{figure}
\centering
\includegraphics[width=\textwidth]{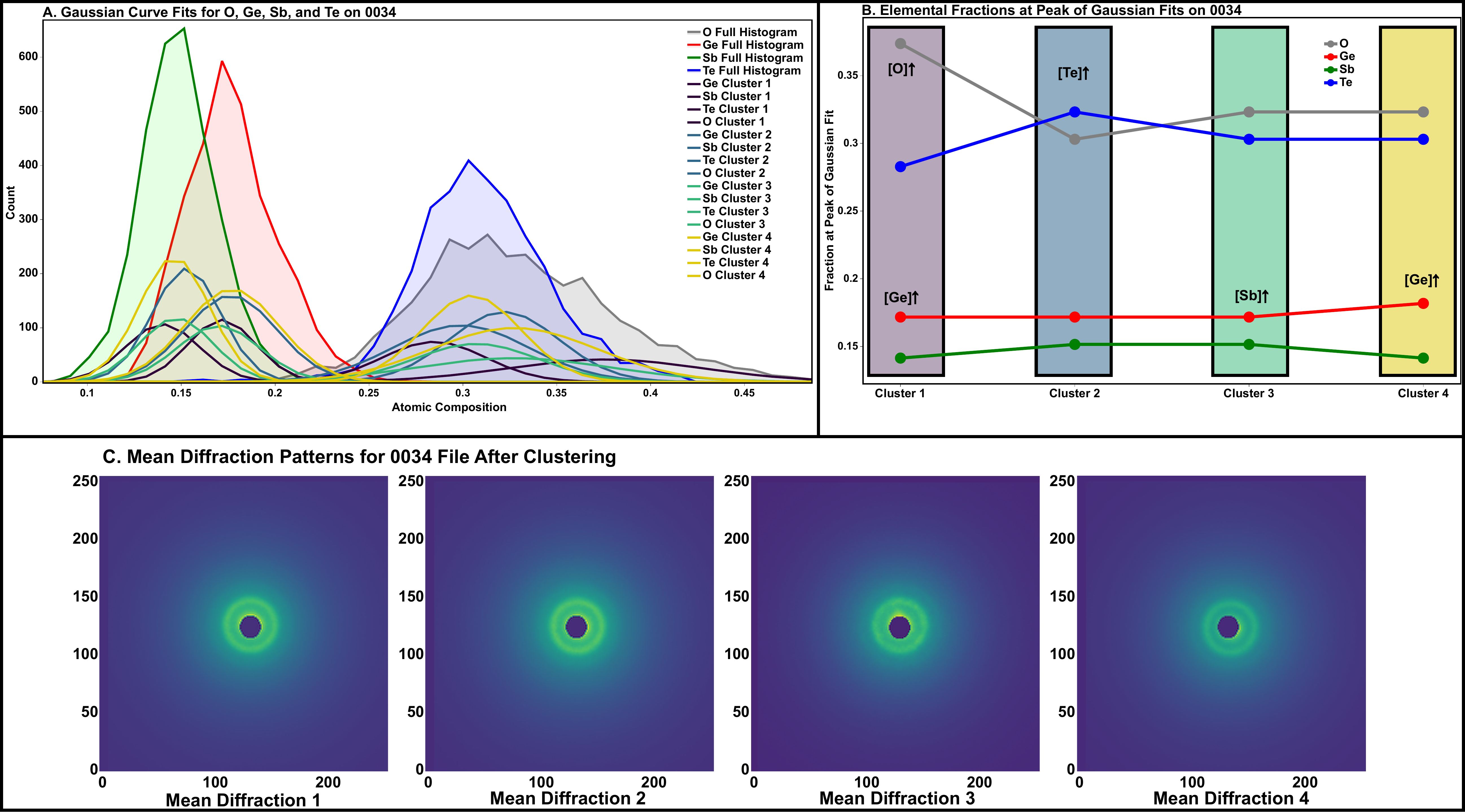}
\caption{A. Application of Gaussian Fit on the intensity histograms of EDS values corresponding to different classes obtained from the k-means clustering. B. Maximum atomic fraction of all the elements in different clusters. C. Mean diffraction patterns from the obtained clusters.}
\label{fig:gaussian}
\end{figure}

The objective of this study is to investigate how variations in diffraction patterns correspond to the different phases present in the sample. Specifically, we aim to determine whether it is possible to identify the spatial distribution of distinct atomic phases within the sample based solely on diffraction patterns, using machine learning as the analytical tool.

Since the goal is to distinguish between different phases, the problem aligns conceptually with a classification task in machine learning. However, due to the absence of explicit phase labels for each diffraction pattern, the problem must instead be formulated as an unsupervised clustering task. In this framework, diffraction patterns with similar characteristics, likely corresponding to similar phases or elemental compositions are expected to be grouped together without prior knowledge of their labels.

In our clustering analysis, regions with high germanium content formed distinct clusters separated from tellurium-rich regions, indicating clear compositional segregation in the reduced feature space. Areas exhibiting low concentrations of all elements also clustered separately, occupying distinct locations in the embedding. These patterns suggest that the dimensionality reduction and subsequent clustering successfully captured meaningful compositional heterogeneity across the sample (see Figure \ref{fig:tsne}). This analysis provides a preliminary validation of our hypothesis that regions with similar features tend to cluster together in the high-dimensional feature space, with compositionally distinct areas forming well separated groups after dimensionality reduction. However, while t-SNE is powerful for capturing local neighbourhood structures, it has several limitations in our context \cite{wattenberg2016use}. First, it is computationally expensive for large datasets, making iterative exploration slow. Second, it poorly preserves the global structure of the data, which limits our ability to interpret relationships between distant clusters. Third, its results are highly sensitive to hyperparameters and random initialization, potentially affecting reproducibility. Moreover, t-SNE cannot directly embed new data points without re-running the algorithm on the entire dataset.

To address these challenges, we turned to Uniform Manifold Approximation and Projection (UMAP). UMAP, like t-SNE, is an unsupervised non-linear dimensionality reduction method, but it scales better to large datasets, preserves both local and global structures more effectively, and provides a parametric option for embedding new samples without recomputing the full embedding \cite{becht2019dimensionality,mcinnes2018umap}. This makes UMAP more suitable for our analysis, especially when dealing with large-scale EDS datasets and exploring structural relationships across multiple regions.

After applying UMAP with three components to the PCA-reduced dataset from the 0034 file, we generated both 3D and 2D projections of the embedding (Figure \ref{fig:umap}A and \ref{fig:umap}B). Compared to t-SNE, UMAP exhibited superior preservation of the global data structure, enabling clearer separation and identification of distinct regions in the sample. The improved global and local structure preservation achieved with UMAP provided an embedding that was more suitable for clustering. Since UMAP projects the high-dimensional feature space into a lower-dimensional representation while retaining meaningful relationships between data points, the resulting embedding serves as an ideal input for clustering algorithms. We therefore applied k-means clustering to the UMAP-transformed data to quantitatively partition the sample into distinct groups, enabling a direct comparison between spatial regions and their underlying feature distributions. The classification obtained from k-means clustering was then projected back onto the spatial coordinates of the original sample, allowing us to visually identify distinct regions or phases emerging within the material.

To ensure that the clustering captured meaningful structure in the data, we used the silhouette score as an evaluation metric. The Silhouette score quantifies how similar each point is to others within its assigned cluster compared to points in other clusters. By calculating the Silhouette score for different values of k in k-means, we systematically evaluated the trade-off between compactness within clusters and separation between them. This analysis revealed that k = 4 yielded the highest average Silhouette score (see Figure \ref{fig:flowgraph}), indicating that four distinct clusters provided the best balance between internal cohesion and external separation. We therefore adopted k = 4 as the optimal number of clusters for the final spatial classification, ensuring that the phase separation observed in the projections is both statistically supported and physically interpretable.

Following dimensionality reduction with UMAP, the resulting low-dimensional representation was subjected to K-means clustering to identify groups of points with similar feature-space characteristics. Each data point in this representation corresponds directly to a pixel in the original sample, allowing the cluster assignments to be mapped back to their exact physical locations in real space (see Figure \ref{fig:kmeans}). By reassigning each pixel with its respective cluster label, we effectively generated a spatial phase map of the sample, where different colours represent distinct clusters. This back-projection transforms abstract clusters in feature space into tangible regions within the sample, making it possible to visually identify spatially contiguous domains that share similar spectral or compositional signatures. The result is a powerful visualization of how statistical patterns in the high-dimensional data manifest as chemically or structurally distinct areas in the material’s physical structure.

In addition to the compositional scatter plots, intensity histograms were generated for each cluster by projecting the corresponding elemental compositions from the EDS maps of O, Ge, Sb, and Te. To enhance interpretability, these histograms were Gaussian-fitted, providing a smooth representation of the intensity distributions and reducing noise-related fluctuations. The fitted curves highlighted the characteristic peaks for each element within a given cluster, making it easier to identify subtle differences in composition between regions (see Figure \ref{fig:gaussian}A). Furthermore, the characteristic peaks of each element in the clusters were plotted to compare their elemental compositions. Our analysis reveals that Cluster 1 is enriched in oxygen and germanium. In Cluster 2, the concentration of tellurium increases, while in Cluster 3, antimony shows a relative increase. In Cluster 4, germanium concentration rises again (see Figure \ref{fig:gaussian}B). Finally, average diffraction patterns were generated from the clustering data in order to provide a direct comparison of the structural characteristics across the different clusters. This approach enables the visualization of subtle variations in diffraction features, thereby offering further insight into the distinct structural signatures associated with each cluster (see Figure \ref{fig:gaussian}C). 

\section*{Conclusion}

In this work, clustering analysis was applied to 4D-STEM diffraction data of the phase-change material Ge-Sb-Te to resolve both compositional and structural heterogeneity. Following data preprocessing to remove artifacts, dimensionality reduction was performed using PCA, with t-SNE and UMAP employed to validate and enhance cluster separation in the embeddings in 3-dimensions. Subsequent k-means clustering, optimized by silhouette score analysis, identified four distinct clusters. Mapping these clusters back to the original data and analyzing intensity histograms revealed clear chemical signatures: oxygen and germanium enrichment in Cluster 1, increased tellurium concentration in Cluster 2, enhanced antimony content in Cluster 3, and elevated germanium levels again in Cluster 4. These compositional distinctions were further supported by averaged diffraction patterns, which underscored structural differences across clusters. Collectively, these results demonstrate that unsupervised clustering provides a robust framework for correlating local elemental composition with structural features in complex material systems. Looking ahead, integrating this approach with advanced spectroscopic and computational methods may further refine cluster resolution, offering new opportunities for the design and optimization of next-generation functional materials.

\section*{Materials and Methods}

\subsection*{Preprocessing of 4DSTEM data}
A single sample of phase changing material (Ge-Sb-Te) was imaged by Thermo Fisher Spectra 200 STEM microscope. The sample was scanned multiple times at a magnification of 185k times to give mixed raster content (MRC) files namely from 0033 to 0038. Each diffraction pattern was $256 \times 256$ and real space dimension of the sample was $110 \times 45$. Other important information can be found in the Table \ref{tab:datainfo} \cite{kumar2025predictionedsmaps4dstem}.

\begin{table}[h]
\setlength{\tabcolsep}{2pt}
\renewcommand{\arraystretch}{1.2}

\caption{This table shows import properties of the data.}
\label{tab:datainfo}
\centering
\begin{tabular}{|C{6cm}|C{6cm}|}
\toprule

\textbf{Feature} & \textbf{Value} \\
\midrule

Spot Size & 2.5 nm \\

R Pixel Size & 1.053 nm \\

Q Pixel Size & 0.1451 $A^{-1}$ \\

Image Bit Depth & 16-bit \\
\bottomrule
\end{tabular}

\end{table}

\begin{figure}[t]
\centering
\includegraphics[width=\textwidth]{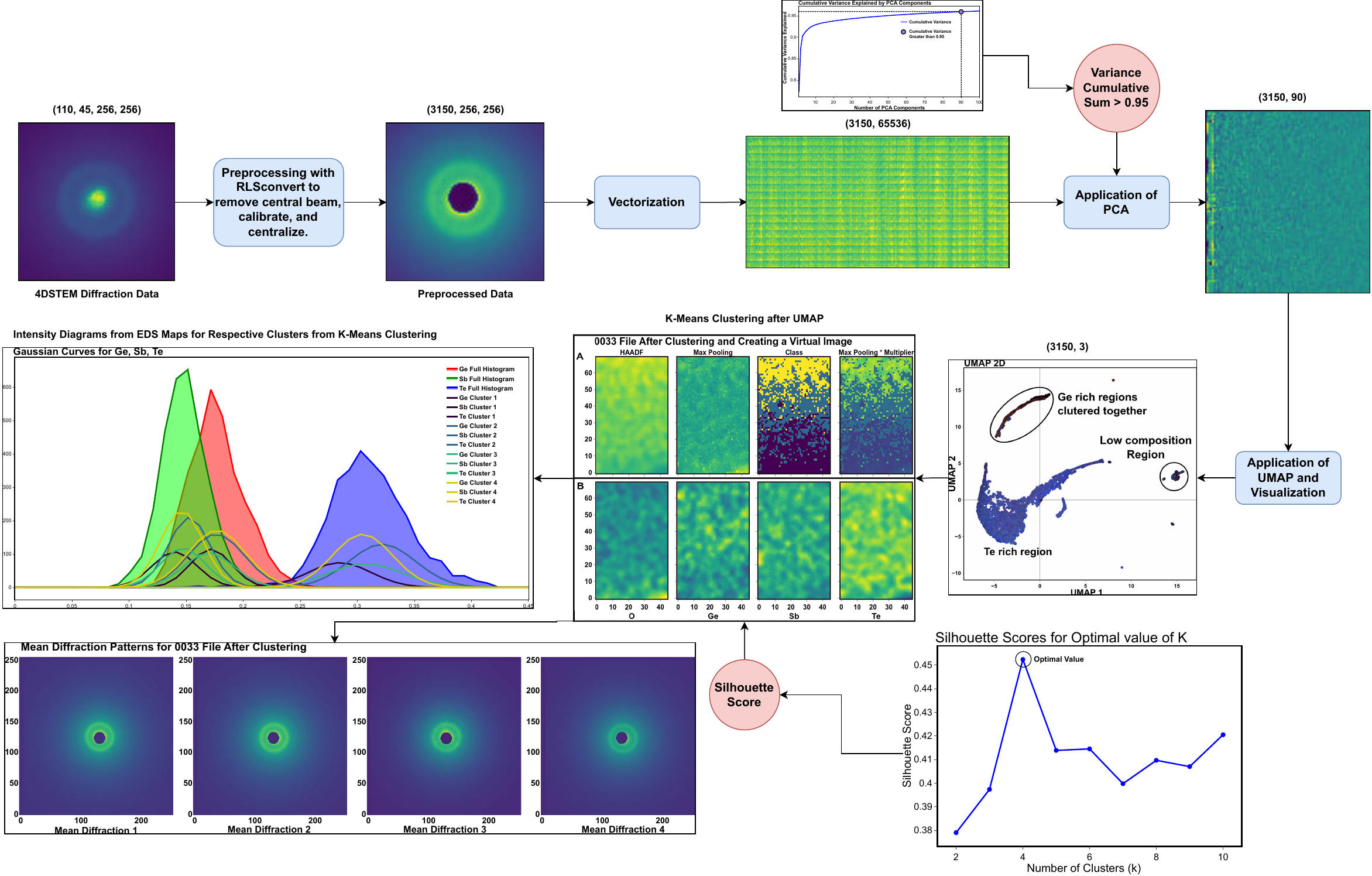}
\caption{This figure shows the various steps involved in training the machine learning model for the determination of different phases in the sample.}
\label{fig:flowgraph}
\end{figure}

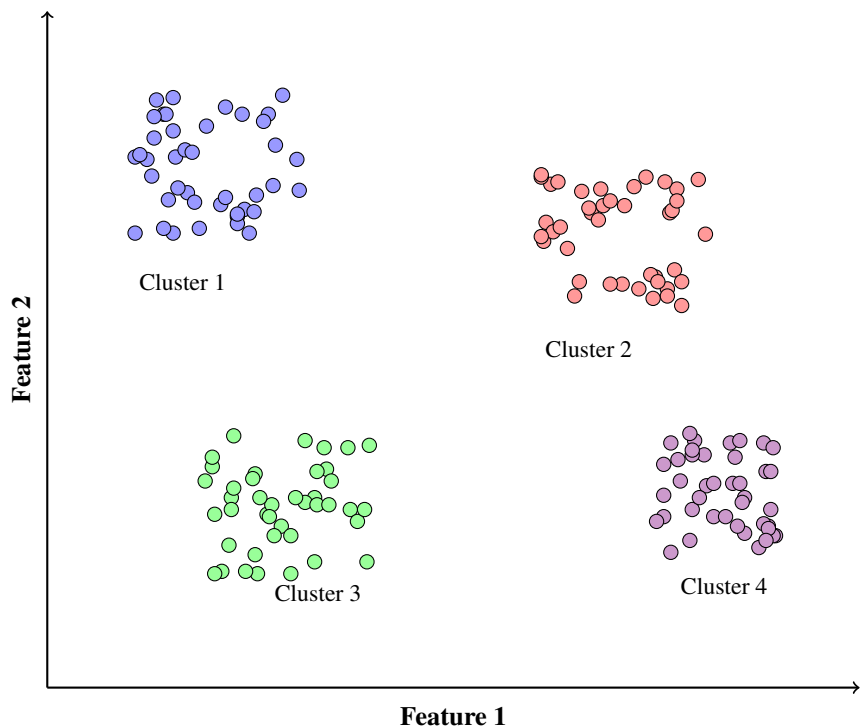
\begin{figure}[t]
  \centering
  \begin{tikzpicture}[scale=0.9]
  \draw [->, thick](0, 0) -- (12, 0) node [anchor = south, align = center] at (6, -0.7) {\textbf{Feature 1}};
  \draw [->, thick] (0, 0) -- (0, 10) node [anchor = south, align = center, rotate = 90] at (-0.1, 5) {\textbf{Feature 2}};
  
  \node [] at (8, 5) {\small Cluster 2};
  \pgfmathrandominteger{\noc}{40}{40}
  \foreach \i in {0,1,...,\noc}{
  	\pgfmathrandominteger{\x}{-20}{50};
  	\pgfmathrandominteger{\y}{-10}{50};
	\draw [fill = red!40] (8cm + \x pt, 6cm + \y pt) circle(3pt);}
	
	\node [] at (2, 6) {\small Cluster 1};	
	\foreach \i in {0,1,...,\noc}{
	  	\pgfmathrandominteger{\x}{-20}{50};
	  	\pgfmathrandominteger{\y}{-10}{50};
		\draw [fill = blue!40] (2cm + \x pt, 7cm + \y pt) circle(3pt);}  
	
	\node [] at (4, 1.4) {\small Cluster 3};
	\foreach \i in {0,1,...,\noc}{
	  	\pgfmathrandominteger{\x}{-20}{50};
	  	\pgfmathrandominteger{\y}{-10}{50};
		\draw [fill = green!40] (3cm + \x pt, 2cm + \y pt) circle(3pt);}
	\node [] at (10, 1.5) {\small Cluster 4};
	\foreach \i in {0,1,...,\noc}{
	  	\pgfmathrandominteger{\x}{0}{50};
	  	\pgfmathrandominteger{\y}{0}{50};
		\draw [fill = violet!40] (9cm + \x pt, 2cm + \y pt) circle(3pt);}
  \end{tikzpicture}
  \caption{This figure shows the collection of diffraction patterns in feature space. Points which are closer to each other are closely related.}
  \label{fig:clustering}
\end{figure}

The raw experimental data were originally stored in MRC format, which encodes three-dimensional diffraction datasets. To enable efficient storage, processing, and analysis, these files were converted into four-dimensional hierarchical data format (HDF5) using the \lstinline|rlsconvert| package. In addition to format conversion, this package provides essential preprocessing utilities, including central beam removal, diffraction pattern calibration, and centralization (see Figure \ref{fig:flowgraph}).

The raw MRC files contained unwanted regions at the top (artefacts) and bottom (highly oxidized area) of the dataset, which were systematically cropped from all samples during preprocessing. The initial dataset size of (4950,256,256) was transformed into  (110,45,256,256) after masking, centralizing, and calibration using \lstinline|rlsconvert|. Following the cropping of the upper and lower regions, the final dataset size was reduced to (70,45,256,256).

For subsequent machine learning analysis, the diffraction patterns were vectorized, reshaping each $256 \times 256$ pattern into a one-dimensional vector. This resulted in a dataset of size (3150,65536). Since the dimensionality of the feature space is very high, applying clustering algorithms directly would be computationally expensive and potentially inefficient. To reduce dimensionality while retaining the essential variance, principal component analysis (PCA) was applied, compressing the feature space from 65536 to 90 components while preserving more than 95\% of the explained variance (see Code \ref{lst:scree_plot}). This dimensionality reduction step ensured computational efficiency.

Since the available phases of the sample are not labelled for specific regions, and only the 4D-STEM diffraction patterns and EDS maps are available, the task can be formulated as an unsupervised clustering problem \cite{wei2019machine}. The basic idea behind the study is that the diffraction patterns having similar features would be closer in the N-dimensional feature space (see Figure \ref{fig:clustering}).

\begin{lstlisting}[language = Python, caption = {This code shows the determination of the optimal number of PCA components.}, label={lst:scree_plot}]
from sklearn.decomposition import PCA
import Py4DSTEM
import numpy as np
X = Py4DSTEM.read("0034.h5").reshape((4950, 65536))
pca = PCA().fit(X)
cumulative_variance = np.cumsum(pca.explained_variance_ratio_)
n_components = np.argmax(cumulative_variance >= 0.95) + 1
print("Number of components for 95% variance:", n_components)
\end{lstlisting}

\subsection*{Application of Clustering Algorithms}
\subsubsection*{Application of t-SNE}
After the preprocessing step, we first applied t-Distributed Stochastic Neighbour Embedding (t-SNE) \cite{maaten2008visualizing} using \lstinline|scikit-learn| (v1.6.1) implementation. The method was run with a perplexity of 30 and auto learning rate. t-SNE enabled the visualization of local similarities among diffraction patterns in a three dimensional embedding space, providing an initial indication of clustering behaviour (see Figure \ref{fig:tsne}). The marker colour were used from the compositional values from EDS maps as red for Ge, green for Sb, and blue for Te. However, consistent with known limitations of t-SNE \cite{wattenberg2016use}, we observed distortions in global geometry and sensitivity to parameter choice, particularly with our large dataset.

\subsubsection*{Application of UMAP}
To address the issues with t-SNE, we subsequently applied Uniform Manifold Approximation and Projection (UMAP) \cite{becht2019dimensionality,mcinnes2018umap} using the umap-learn (v0.5) package. UMAP was configured with k nearest neighbours and a minimum distance parameter of 0.1, projecting the high-dimensional patterns into both 2D and 3D manifold. Each marker in obtained embedding was assigned a colour similar to t-SNE clusters. Compared to t-SNE, UMAP preserved both local and global relationships more effectively and scaled more efficiently to our dataset size (see Figure \ref{fig:umap}).

\subsubsection*{Application of k-Means Clustering}
Following dimensionality reduction, clustering was performed using the k-means algorithm \cite{ostrovsky2013effectiveness}. The reduced embeddings obtained from t-SNE and UMAP served as the input feature space, allowing the algorithm to partition the data into k distinct clusters corresponding to regions with similar diffraction characteristics. The number of clusters was determined by evaluating the inertia and silhouette scores across different k values, ensuring stable and physically interpretable groupings. This approach has been widely adopted in microscopy and diffraction analysis for segmenting high-dimensional data into meaningful structural classes \cite{wei2019machine}.

After applying k-means clustering, the assigned cluster labels were mapped back onto the spatial coordinates of the original 4D-STEM dataset. This projection enabled the visualization of cluster membership directly onto the sample, allowing each pixel or probe position to be colour-coded according to its cluster class. In this way, regions with similar diffraction patterns could be identified and correlated with the corresponding EDS maps, facilitating the interpretation of structural and compositional variations across the sample.

\end{document}